\documentclass[fleqn,10pt]{wlscirep}
\usepackage[utf8]{inputenc}
\usepackage[T1]{fontenc}
\title{Gravity-Assist as a Solution to Save Earth from Global Warming}

\author[1,*]{Sohrab Rahvar}
\affil[1]{Physics Department, Sharif University\\
P.O.Box 11365-9161, Azadi Avenue, Tehran,Iran}

\affil[*]{rahvar@sharif.edu}

\begin{abstract}
Global warming is one of the problems of human civilization and decarbonization policy is the main solution to this problem. In this work, we propose an alternative method of using the gravity-assist by the asteroids to increase the orbital distance of the Earth from the Sun. We can manipulate the orbit of asteroids in the asteroid belt by solar sailing and propulsion engines to guide them towards the Mars orbit and a gravitational scattering can put asteroids in a favorable direction to provide an energy loss scattering from the Earth. The result would be increasing the orbital distance of the earth and consequently cooling down the Earth's temperature. We calculate the increase in the orbital distance of the earth for each scattering and investigate the feasibility of performing this project.   
\end{abstract}
\begin{document}
\flushbottom
\maketitle 

%\begin{keywords}
%\end{keywords}

\section{Introduction}

\begin{figure}
\centering
\includegraphics[scale=0.36]{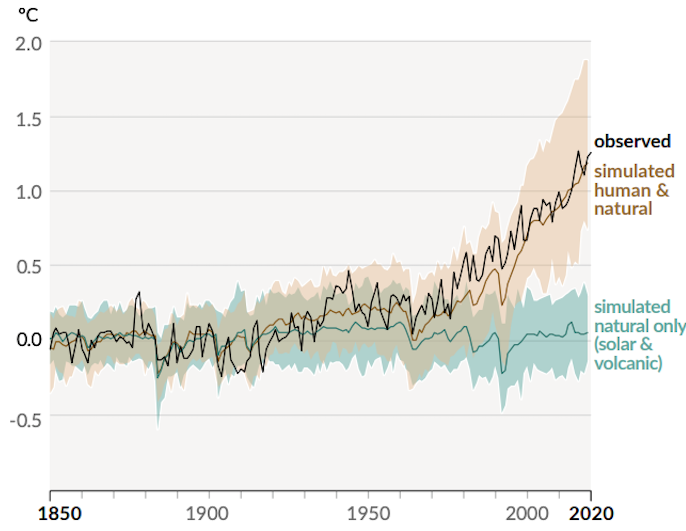}
%\vspace{-1cm}
\caption{ Change in global surface temperature (annual average) as observed and
simulated using human \& natural and only natural factors (both 1850-2020), adapted from \cite{allan} }
\label{fig1}
\end{figure}

The monitoring of the Earth's climate shows a significant change in temperature and climate pattern due to the emission of greenhouse gases by humans \cite{pac}.  The two gases of carbon dioxide and methane as a result of fuel burning are the main components of the greenhouse gases causing the climate change. In recent years the temperature rise is accelerated as a result of feedback from the melting of ices and decreasing of the reflection and increasing of the water vapor in the atmosphere \cite{Jia}.  Figure (\ref{fig1}) shows the trend of temperature increase in terms of the year which compares the measured temperature of earth with the simulations  \cite{allan}.  In recent years the increase in extreme events of weather also shows a direct correlation with the increase of the temperature of the atmosphere and oceans \cite{ext}. 

The present policy to prevent  the global temperature is mainly based on decreasing fossil fuel consumption and investment in green sources of energy. The policy of decarbonization is based on the rapid development of renewable energies \cite{rene}. The Paris agreement as an international implementation plans that the mean global temperature doesn't rise beyond $2^\circ$ C above pre-industrial levels.

In this work, we propose an alternative method for decreasing the global temperature by changing the orbital distance of Earth from the Sun. This operation can be done by the gravity-assist mechanism with the flyby motion of the asteroids close to the earth where the energy loss of asteroids can increase the orbital distance of the Earth from the Sun and results in a decrease in the Earth temperature. While this plan may look like an out-of-practical idea, we will show the feasibility of this project at least for future technology. 

In Section (\ref{sec1}), we introduce the gravity-assist and provide the physical mechanism to use the asteroids to 
change the mechanical energy of the Earth. In section (\ref{temp}) we discuss the dependence of mechanical energy of the Earth to the global temperature and calculate the temperature change on earth for an asteroid scattering with a given mass. In Section (\ref{sc}), we provide a practical plan for the asteroid scattering from the earth. The Discussion is given in (\ref{con}). 

\section{Gravity-assist and energy loss process}
\label{sec1}
The gravity-assist is suggested by Yuri Kondratyuk in 1938 for accelerating a spacecraft traveling between the planets of the solar system using the gravity of planets as the slingshot \cite{yuri}.  In gravity assist the relative speed of the spacecraft before and after the gravitational scattering with respect to a planet is the same, however since the planet is moving with respect to the sun, the spacecraft can accelerate or decelerate with respect to the solar frame.  The gravity assist maneuver was first used in 1959 by the Soviet probe Luna 3 for monitoring the Moon. The other interplanetary probes such as the Pioneer program, Mariner 10, Voyager program, Galileo, Ulysses, and many other programs used gravity-assist to reach the other planets at the edge of the solar system. The gravity assist not only can change the kinetic energy of the spacecraft also can change the energy of photons or the frequency of light during the gravitational scattering so-called gravitational lensing \cite{rahvar}. 

\begin{figure}
\centering
\includegraphics[scale=0.20]{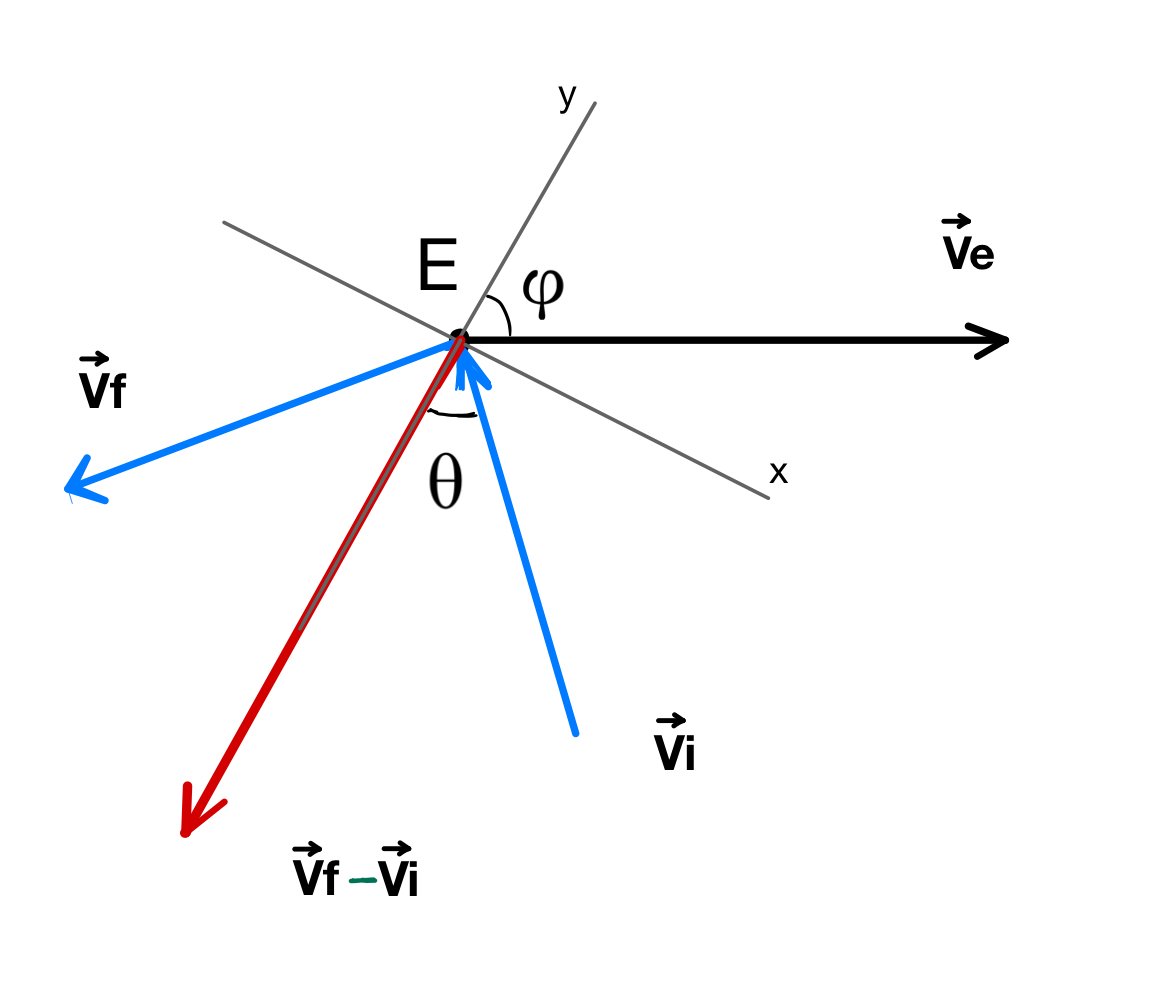}
\caption{ Schematic figure for the scattering where in this case $\Delta E_e >0$ and earth can gain energy from this scattering. $\bf{v_e}$ (with black arrow) is the velocity of earth.  $\bf{v}_i$ and
 $\bf{v}_f$ (in blue arrows) are the relative velocity of asteroid with respect to the earth before and after the scattering. The red arrow is the difference of $\bf{v}_f - \bf{v}_i$ and the grey axis are the symmetric coordinate system with respect to $\bf{v}_i$  and $\bf{v}_f$. 
%axis represents along the symmetric axis with respect to $\vec{v}_i$ and
 %$\vec{v}_f$ and $\phi$ and $\theta$ represent angles of $\vec{v}_e$ and $\vec{v}_i$ with respect to x-axis.
 }
\label{fig2}
\end{figure}

Here we want to use the gravity-assist mechanism for the scattering of asteroids from the Earth and instead of energy-gain by the asteroid, we manage the energy loss of asteroid where in this case the energy of asteroid transfers to the mechanical energy of Earth, orbiting around the Sun. The result of this scattering would be increasing the orbital distance from the sun and consequently cooling down the temperature of the planet.

%To investigate the feasibility of this project we categorized asteroids based on their orbiting positions in the solar system as \textcolor{blue}{(i) the asteroid belt, (ii) Trojans, and (iii) Near-Earth asteroids \citep{locate,trojan,ne}. We will discuss each category in detail.}   

In the following, we investigate the scattering process of an asteroid from Earth. Let us take $m_a$ as the mass of an asteroid and $M_e$ as the mass of the earth. ${\bf v}_i$ and ${\bf v}_f$ are the initial and final velocity vectors of the asteroid  (as a result of scattering) with the respect to the Earth and ${\bf v'}_i$ and ${\bf v'}_f$ are the same velocities with respect to the Sun. Assuming that the mass of the asteroid is much smaller than earth (i.e. $m_a\ll M_e$), the speed of the asteroid before and after the scattering  with respect to the Earth is equal (i.e. $|{\bf v}_i| = |{\bf v}_f|$). However the velocity of asteroid with respect to the sun before the scattering is $\bf{v'}_i = \bf{v}_i + \bf{v}_e$ and after the scattering is ${\bf v'}_f = {\bf v}_f + {\bf v}_e$. Here ${\bf v}_e$ is the velocity of the earth with respect to the sun. 
We note that the incoming and outgoing speed of asteroid with respect to the solar coordinate are not equal.

The energy difference of asteroid after the scattering compare to that before the scattering in the frame of sun is given by  
$\Delta E_a = m_a {\bf v}_e\cdot({\bf v}_f - \bf{v}_i)$. According to the conservation of energy, the energy gain of earth would be $\Delta E_e = - {\bf v}_e\cdot({\bf p}_{f} - {\bf p}_{i})$ where ${\bf p}_{f} = m_a {\bf v}_f$ and ${\bf p}_{i} = m_a {\bf v}_i$ are the momentum of asteroid with respect to the Earth frame where $| {\bf p}_{i} | = | {\bf p}_{f} |$.  Also the momentum of Earth would change by the amount of $\Delta {\bf P}_e = - ({\bf p}_{f} - {\bf p}_{i})$. In order to calculate the energy and momentum change of earth, we project the velocities on $x$ and $y$ axis in Fig (\ref{fig2}) as follows:
\begin{eqnarray}
{\bf v}_e = v_e(\hat{i} \sin\phi + \hat{j} \cos\phi),\\
{\bf v}_i  = v_i (- \hat{i} \sin\theta + \hat{j} \cos\theta),\\
{\bf v}_f  = v_i (- \hat{i} \sin\theta - \hat{j} \cos\theta),
\end{eqnarray}
where the $y$ axis of this coordinate system is designed to be symmetric with respect to the incoming and outgoing velocities 
of the asteroid with respect to the earth. The $\theta$ angle is the direction of incoming velocity with the $y$-axis and the $\phi$ angle is earth's velocity 
with the $y$-axis. Substituting in the energy and momentum gain of earth results in 
\begin{eqnarray}
\Delta E_e &=& 2m_a v_i v_e \cos\theta \cos\phi, \\
\Delta {\bf p_e} &=& 2 m_a v_i \cos\theta~  {\hat j}.
\end{eqnarray}
Using the total mechanical energy of earth around the sun $E_e = -{GM_\odot M_e}/{2R_e}$, then the relative energy change of Earth orbiting around the sun would be 
\begin{equation}
\label{ediff}
\frac{\Delta E_e}{|E_e|} = 4(\frac{m_a}{M_e}) (\frac{v_i}{v_e}) \cos\theta \cos\phi .
\end{equation}
\section{Temperature change of earth as a result mechanical energy change} 
\label{temp}
The change in the total energy of earth results in the change in the orbital distance as 
$\Delta E/|E| = \Delta R/R$. Since the power of light received by the earth depends to distance as $S \propto 1/R^2$, 
 we can conclude that $\Delta S/S_0 = -2\Delta E_e/|E_e|$. 
Then from equation (\ref{ediff}), the relative change in the power of sun received  by the earth would be 
\begin{equation}
\frac{\Delta S}{S_0} = -0.5\times 10^{-9} (\frac{m_a}{10^{15} \text{kg}})(\frac{v_i}{11 \text{km/s}})\cos\theta~\cos\phi . 
\label{ds}
\end{equation}
The change in the flux of light received by the earth can change the average global temperature. While this is a complicated process, however, we can have a rough estimation on the temperature decrease.

The spectrum of the sun's light that is absorbed by the earth is reemitted in the longer wavelengths as the blackbody radiation. The total rate of energy that is absorbed by the earth is $\dot{\cal{E}}_{in} = S_0 \pi r_e^2 (1-A_e)$ where $S_0$ is the energy flux of the sun at the position of the earth, $\pi r_e^2$ is the effective area of earth and $A_e$ is the Albedo of earth. On the other hand earth radiates as a blackbody in the longer wavelength with the rate of $\dot{\cal{E}}_{out} = 4\pi r_e^2\sigma T_e^4$. For the equilibrium condition of energy between the inward and outward energy flux, the effective temperature of the earth would be
\begin{equation}
T_e = \left(\frac{S_0 (1-A_e)}{4\sigma}\right)^{1/4},
\end{equation}
where for the earth albedo of $A\simeq 0.3$ \cite{et}, the effective temperature of earth would be $T_e = 252$K. This temperature is smaller than the average ground temperature of the earth. However, using the greenhouse effect where the atmosphere can absorb part of energy radiating from the surface of earth the temperature of the ground would be 
\begin{equation}
T_g =  \left(\frac{S_0 (1-A_e)}{2\sigma (2-\epsilon)}\right)^{1/4},
\label{T}
\end{equation}
where $\epsilon$ is the effective  emissivity of ground. We adapt $\epsilon \simeq 0.9$.  
Then the temperature on ground of earth enhances to $T_g\sim 300$K.  Substituting equation (\ref{ds}) in the derivative of equation (\ref{T}) results in the temperature change on the ground of earth due to orbital 
change of earth as
 \begin{equation}\label{dt}\Delta T_g = - 0.35\times 10^{-7}~K (\frac{T_g}{300 \text{K}})(\frac{m_a}{10^{15} \text{kg}})(\frac{v_i}{11 \text{km/s}})\cos\theta~\cos\phi.
\end{equation}

We note that the deflection angle of the asteroid scattering by the earth (i.e. $\theta$) depends on the impact parameter (as the minimum distance of asteroid from the earth) as well as the initial velocity of the asteroid. For an asteroid with positive energy of $E$ with respect to the earth, the shape of orbit in polar coordinate is a hyperbola with the equation of $\alpha/r = 1 + \epsilon\cos\gamma$. Here, $\epsilon$ is the eccentricity and depends on the physical parameters of asteroid as \cite{marion} 
\begin{equation}
\epsilon = (1 + \frac{2EL^2}{\mu \kappa^2})^{1/2},
\end{equation}
where $L$ is the angular momentum of the asteroid with respect to the earth, $\kappa = GM_em_a$ and $\mu$ is the reduced mass of the asteroid and since $\mu\ll M_e$ then $\mu\simeq m_a$. The scattering angle is determined from $\epsilon$ (for $r\rightarrow \infty$) as $\cos\gamma = -1/\epsilon$ where $\gamma$ depends in $\theta$ in Fig. (\ref{fig2}) by $\gamma = \pi - \theta$, then $\cos\theta = 1/\epsilon$.

Using $L=m_av_i b$ where $b$ is the impact factor, the eccentricity can be written as $\epsilon = (1 + m_a^2 v_i^2 b^2/\kappa^2)^{1/2}$. We replace $\cos\theta=1/\epsilon$ in equation (\ref{dt}) in terms of the eccentricity parameter and in terms of the escape velocity from the earth and the impact parameter, where equation (\ref{dt}) simplifies to  
\begin{equation}
\Delta T_g = - 0.35\times 10^{-7}~K (\frac{T_e}{300 \text{K}})(\frac{m_a}{10^{15} \text{kg}})(\frac{v_i}{11 \text{km/s}})
\left(1+(\frac{v_i}{11\text{km/s}})^2(\frac{b}{6400~\text{km}})^2\right)^{-1/2}\cos\phi. 
\label{dt22}
\end{equation}
Here we normalized the initial velocity of the asteroid to the escape velocity from the earth and the impact parameter to the radius of the earth.

The distribution of the asteroids in terms of their size in the main belt of asteroids within the range of $2.82~ \text{a.u.} <a< 2.96~ \text{a.u.}$ is a power law function as $N(D) \sim D^q$ where $q$ is $-2.5$ for 
the range of $100~\text{km} < D < 1000~\text{km}$ and a shallower slope for $D < 100$~km, the exponent is $q \sim -1.8$ down to $D \sim 10$~km \cite{dist}. Let us assume an average density of $\bar{\rho} = 2$ gr/cm$^3$ for the asteroids  \cite{density}. As a result, we have more low mass asteroids compare to the large ones. Using the mass of asteroid in terms of the size  
\begin{equation}
m_a(R) =  (\frac{L}{10 \text{km}})^3 \times 10^{15}~  \text{kg},
\end{equation}
then equation (\ref{dt22}) is rewritten as  
\begin{equation}
\label{dt2}
\Delta T_g = - 0.35\times 10^{-7}~K (\frac{T_e}{300 \text{K}})(\frac{L}{10 \text{km}})^3(\frac{v_i}{11 \text{km/s}})
\left(1+(\frac{v_i}{11km/s})^2(\frac{b}{6400~km})^2\right)^{-1/2}\cos\phi. 
\end{equation}

In the rest of this paper, we will discuss how to manage the asteroids  to have a gravitational scattered from the earth. 
\section{Asteroid scattering from the Earth}
\label{sc}
Most of the asteroids in the plane of solar system are moving  in the same direction as the planets move. Our aim is to lower the orbit of the asteroids and encounter them gravitationally with the earth. However, we note that in order to have energy gain for the earth, asteroids should have a head-on gravitational scattering with the earth. 

The closest-large number of asteroids to earth are located in the asteroid belt and in order to have gravitational interaction of asteroids with the earth, we manage to reduce their orbital distance with respect to the sun. The least costing method might be using solar sailing. Also, we can consider  a hybrid braking system using solar sailing and install propulsion engines on the asteroids. For solar sailing, the momentum transfer from the photons of the sun to the asteroid plays the role of braking along with the orbital velocity and the result would be a spiral motion of asteroids towards the lower orbits. 

\begin{figure}
 \centering
\includegraphics[scale=0.25]{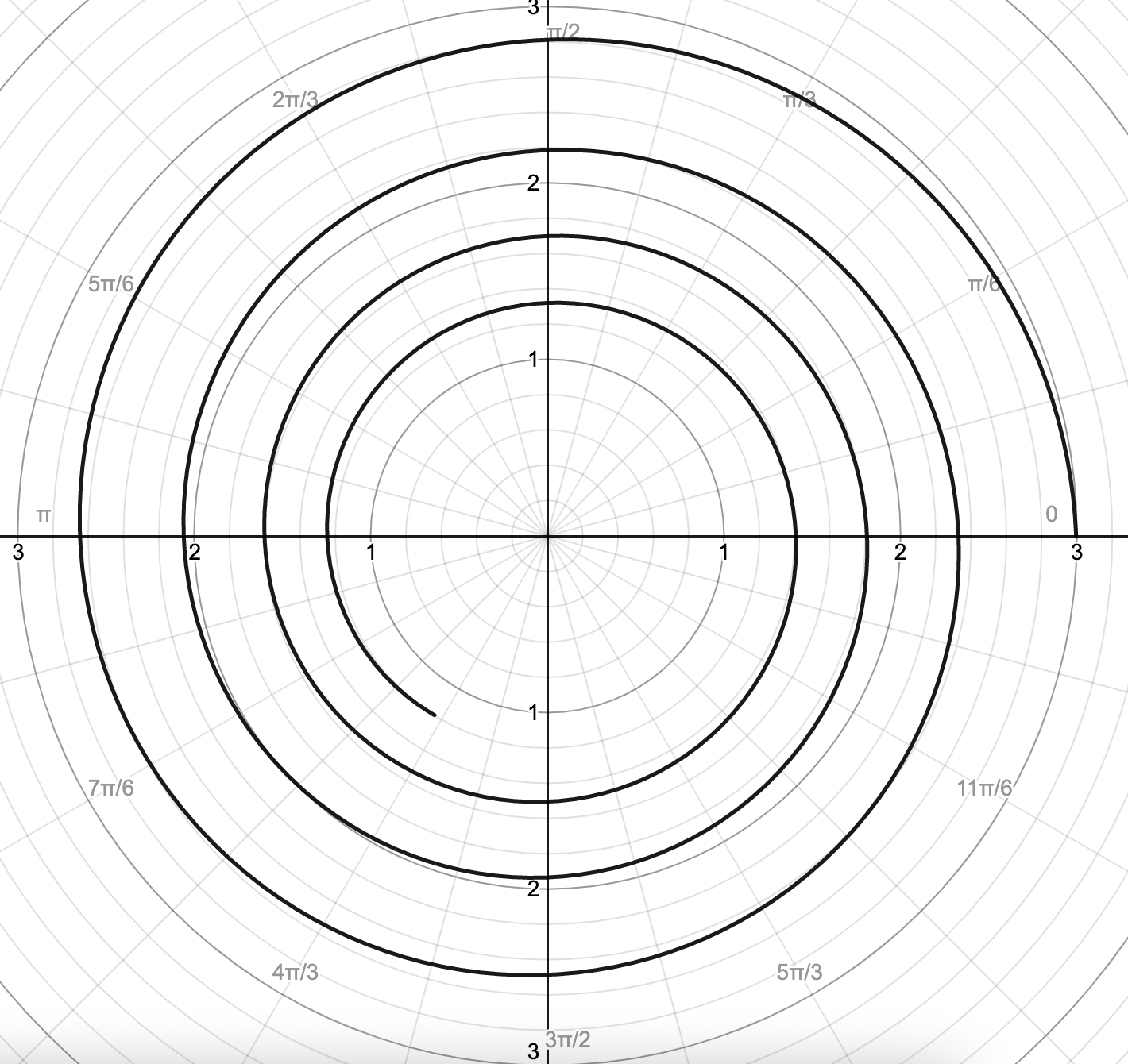}
\caption{Schematic path of an asteroid with the equation of motion provided by $r = r_0 \exp(\beta\varphi)$ with $\beta<0$. %The asteroid 
%starts its decelerating journey from the asteroid belt and approach to the Mars orbit.
}
\label{fig3}
\end{figure}

 Let us denote $r$ as the distance of the sun to the asteroid and $\varphi$ as the polar coordinate and $r$ and the tangent line to the solar sail installed on the asteroid. The general solution for an object with the solar pressure results in 
 the solution of $r = r_0 \exp(\beta\varphi)$ where $r_0$ is the initial orbital distance and $\beta$ is a constant where depending on the sign of it orbital distance changes as shown in Figure (\ref{fig3}). In Methods section, we provide the detailed dynamical solution for the asteroid under radiation pressure.

 The time scale for changing the orbital distance of asteroid can be obtained from $t_c = r/\dot{r}$ where substituting the dynamics of spiral motion of asteroid we obtain  
 \begin{equation} 
 t_c = 74 \text{yr}~ (\frac{m_a}{10^{10}kg})(\frac{v_a}{10 km/s})(\frac{A}{100 km^2})^{-1},
 \end{equation}  
 where $m_a$ is the mass of asteroid, $v_a$ is the orbital velocity of asteroid and $A$ is the area of solar sail installed on the asteroid. In this calculation, we set the solar pressure at the position of sun, $P_0 = S_0/c \simeq 4.5~ \mu$pa. Installing the jet propulsion engines on the asteroids can boost  brakes and provides a shorter time scales for changing the orbit of asteroid. If we lower the orbit of the asteroids to the position of Mars, then a controlled scattering of the asteroid from Mars can guide it toward the earth through an elliptical path. We note that the asteroid and Mars are moving in the same direction and by putting the asteroid in a Hohmann orbit (using the gravity assist) we can have the proper configuration of asteroid-earth scattering as Fig (\ref{fig2}). The result would be an energy loss of the asteroid and enhancing the mechanical energy of the earth. 
 
% change the orbit of planet while asteroid and Mars are going to the same direction, to an elliptical shape with the smaller distance smaller than the orbital radius of earth. In this case we can have configuration of scattering as Figure ({fig2}) where an energy loss of asteroid can enhance the mechanical energy of the sun. The energy gain of earth puts the earth in an Hohmann orbit with a smaller increase in the semi-major axis. By regular scattering the asteroids we can  correct this elliptical orbital orbit a circular orbit. 
 
 \section{Discussion}
 \label{con}

Concluding this work, we have introduced the hazards of global warming and besides decarbonization policy, we proposed a mechanical method to change the orbit of the earth by an energy-loss scattering of asteroids from the earth. We proposed the solar-sail as the braking tool to decrease the orbit of the asteroids from the asteroid belt orbit to the Mars orbit. The time scale to lower the orbit is about $70$ yrs for a $10^{10}$ kg mass asteroid. Using the installed propulsion jet engines on the asteroids will decrease this time scale and enable us to do the asteroid maneuvering for a larger number of asteroids. This project can enable us to change the earth's orbit and cool down its temperature by decreasing the energy flux of the sun received by the earth. This project could be feasible for the future technology on earth also habitable planets with intelligent life on it might used it to save their planets while global warming or when their parent star started to inflate to a giant star.

\section*{Materials and Methods}
The dynamics of an asteroid with a solar sail (in the polar coordinate) is given by the following equations 
 \begin{eqnarray}
  \label{d1}
 \ddot{r} - r\dot\varphi^2 &=& -\frac{k}{r^2}+\frac{D^{(0)}_r}{m r^2}\cos\psi, \\
 r\ddot{\varphi} + 2\dot{r}\dot{\varphi} &=&  \frac{D^{(0)}_r}{m r^2}\sin\psi,
 \label{d2}
 \end{eqnarray}
 where $(r,\varphi)$ are the polar coordinates, $k = GM_\odot$ and $-k/r^2$ is the gravitational acceleration due to Sun and $D^{(0)}_r/(mr^2)$ is the extra drag term as the result of the pressure of solar radiation on the sail installed on the asteroid,  $\psi$ angle is the angle between the $r$ vector and a normal vector to the sail (e.g. $\psi =\pi/2$ means the sail is aligned perpendicular to the $r$ direction).   Assuming $P_0$ as the radiation pressure at the position of the earth, then $D^{(0)}_r = P_0 r_0^2 A\cos\psi$ where $r_0$ is the earth-sun distance and $A$ is the area of the sail.  ${D^{(0)}_r}/(mr^2)\times \cos\psi$ is the extra acceleration in the radial direction and ${D^{(0)}_r}/(mr^2)\times \sin\psi$ is the acceleration along the tangent direction. For $0<\psi<\pi/2$ the solar pressure is accelerating and and results increase in the angular momentum of asteroid and causes the increase of the orbital distance and for $-\pi/2<\psi<0$ the effect would be decreasing the angular momentum of the asteroid and hence decreasing the orbital distance to the sun.

 Substituting the suggested solution of $r = r_0 \exp(\beta\varphi)$ in the equations (\ref{d1}) and (\ref{d2}) , we will have a modified Kepler law as $r^3{\dot\varphi}^2 = B(\psi)$ where 
 \begin{equation}
 B(\psi) = \frac{k}{1+\beta^2} - \frac{D^{(0)}_r  (\cos\psi - \beta\sin\psi)}{m(1+\beta^2)}.
 \label{A}
 \end{equation}
Also the dynamics of asteroid as a function of time obtain as 
\begin{equation}
    r(t) = \left( r_0^{3/2} + \frac{3\beta}{2}\sqrt{B(\psi)}~ t\right)^{2/3},
\end{equation}
where depending on the sign of $\beta$, the orbit of asteroid increase or decrease. 
 We substitute $r=r_0\exp(\beta\varphi)$ in the dynamical equations (\ref{d1} and \ref{d2}) and use the equation (\ref{A}) to  obtain $\beta$ in terms of the gravitational force and radiation pressure as
 \begin{equation}
 \beta = \frac{P_0  r_0^2 A\sin2\psi}{km-P_0 r_0^2 A \cos^2\psi}\simeq \frac{P_0 r_0^2 A}{km}\sin2\psi .
 \end{equation}
  For $-\pi/2<\psi<0$ the braking effect results in a spiral motion of the asteroid towards the lower orbit and $\psi = -\pi/4$  provides the maximum braking. Figure (\ref{fig3}) represents the schematic trajectory of the asteroid towards the lower orbit. We note that our result for the spiral dynamics of the object under solar pressure is different than that is reported in \cite{Ralph}.

{\bf Acknowledgment}
This research
was supported by Sharif University of Technology’s Office of Vice
President for Research under Grant No. G950214

{\bf Data Availability:}
 No new data were generated or analysed in support of this research.

\bibliography{references}

\begin{thebibliography}{10}
\urlstyle{rm}
\expandafter\ifx\csname url\endcsname\relax
  \def\url#1{\texttt{#1}}\fi
\expandafter\ifx\csname urlprefix\endcsname\relax\def\urlprefix{URL }\fi
\expandafter\ifx\csname doiprefix\endcsname\relax\def\doiprefix{DOI: }\fi
\providecommand{\bibinfo}[2]{#2}
\providecommand{\eprint}[2][]{\url{#2}}

\bibitem{pac}
\bibinfo{author}{{Jenkins}, S.}, \bibinfo{author}{{Millar}, R.~J.},
  \bibinfo{author}{{Leach}, N.} \& \bibinfo{author}{{Allen}, M.~R.}
\newblock \bibinfo{journal}{\bibinfo{title}{{Framing Climate Goals in Terms of
  Cumulative CO$_{2}$-Forcing-Equivalent Emissions}}}.
\newblock {\emph{\JournalTitle{Geophysical Research Letters}}}
  \textbf{\bibinfo{volume}{45}}, \bibinfo{pages}{2795--2804},
  \doiprefix\url{10.1002/2017GL076173} (\bibinfo{year}{2018}).

\bibitem{Jia}
\bibinfo{author}{{Held }, I.~M.} \& \bibinfo{author}{{ et al.}}
\newblock \bibinfo{journal}{\bibinfo{title}{{Structure and Performance of
  GFDL's CM4.0 Climate Model}}}.
\newblock {\emph{\JournalTitle{Journal of Advances in Modeling Earth Systems}}}
  \textbf{\bibinfo{volume}{11}}, \bibinfo{pages}{3691--3727},
  \doiprefix\url{10.1029/2019MS001829} (\bibinfo{year}{2019}).

\bibitem{allan}
\bibinfo{author}{{Ades}, M.} \& \bibinfo{author}{{ et al.}}
\newblock \bibinfo{journal}{\bibinfo{title}{{Global Climate}}}.
\newblock {\emph{\JournalTitle{Bulletin of the American Meteorological
  Society}}} \textbf{\bibinfo{volume}{101}}, \bibinfo{pages}{S9--S128},
  \doiprefix\url{10.1175/BAMS-D-20-0104.1} (\bibinfo{year}{2020}).

\bibitem{ext}
\bibinfo{author}{{Velden}, C.}, \bibinfo{author}{{Olander}, T.},
  \bibinfo{author}{{Herndon}, D.} \& \bibinfo{author}{{Kossin}, J.~P.}
\newblock \bibinfo{journal}{\bibinfo{title}{{Reprocessing the Most Intense
  Historical Tropical Cyclones in the Satellite Era Using the Advanced Dvorak
  Technique}}}.
\newblock {\emph{\JournalTitle{Monthly Weather Review}}}
  \textbf{\bibinfo{volume}{145}}, \bibinfo{pages}{971--983},
  \doiprefix\url{10.1175/MWR-D-16-0312.1} (\bibinfo{year}{2017}).

\bibitem{rene}
\bibinfo{author}{{Barton}, J.~P.} \& \bibinfo{author}{{Infield}, D.~G.}
\newblock \bibinfo{journal}{\bibinfo{title}{{Energy Storage and Its Use With
  Intermittent Renewable Energy}}}.
\newblock {\emph{\JournalTitle{IEEE Transactions on Energy Conversion}}}
  \textbf{\bibinfo{volume}{19}}, \bibinfo{pages}{441--448},
  \doiprefix\url{10.1109/TEC.2003.822305} (\bibinfo{year}{2004}).

\bibitem{yuri}
\bibinfo{author}{{Jenkins}, S.}, \bibinfo{author}{{Millar}, R.~J.},
  \bibinfo{author}{{Leach}, N.} \& \bibinfo{author}{{Allen}, M.~R.}
\newblock \bibinfo{journal}{\bibinfo{title}{{Framing Climate Goals in Terms of
  Cumulative CO$_{2}$-Forcing-Equivalent Emissions}}}.
\newblock {\emph{\JournalTitle{Geophysical Research Letters}}}
  \textbf{\bibinfo{volume}{45}}, \bibinfo{pages}{2795--2804},
  \doiprefix\url{10.1002/2017GL076173} (\bibinfo{year}{2018}).

\bibitem{rahvar}
\bibinfo{author}{{Rahvar}, S.}
\newblock \bibinfo{journal}{\bibinfo{title}{{Frequency-shift in the
  gravitational microlensing}}}.
\newblock {\emph{\JournalTitle{Phys Rev D}}} \textbf{\bibinfo{volume}{101}},
  \bibinfo{pages}{024015}, \doiprefix\url{10.1103/PhysRevD.101.024015}
  (\bibinfo{year}{2020}).
\newblock \eprint{1908.01361}.

\bibitem{et}
\bibinfo{author}{{Susskind}, J.}, \bibinfo{author}{{Schmidt}, G.~A.},
  \bibinfo{author}{{Lee}, J.~N.} \& \bibinfo{author}{{Iredell}, L.}
\newblock \bibinfo{journal}{\bibinfo{title}{{Recent global warming as confirmed
  by AIRS}}}.
\newblock {\emph{\JournalTitle{Environmental Research Letters}}}
  \textbf{\bibinfo{volume}{14}}, \bibinfo{pages}{044030},
  \doiprefix\url{10.1088/1748-9326/aafd4e} (\bibinfo{year}{2019}).

\bibitem{marion}
\bibinfo{author}{Thornton, S.~T.} \& \bibinfo{author}{{Marion}, J.~B.}
\newblock \emph{\bibinfo{title}{Classical dynamics of particles and systems}}
  (\bibinfo{publisher}{Fort Worth: Saunders College Pub.},
  \bibinfo{year}{1995}).

\bibitem{dist}
\bibinfo{author}{{Tsirvoulis}, G.}, \bibinfo{author}{{Morbidelli}, A.},
  \bibinfo{author}{{Delbo}, M.} \& \bibinfo{author}{{Tsiganis}, K.}
\newblock \bibinfo{journal}{\bibinfo{title}{{Reconstructing the size
  distribution of the primordial Main Belt}}}.
\newblock {\emph{\JournalTitle{Icarus}}} \textbf{\bibinfo{volume}{304}},
  \bibinfo{pages}{14--23}, \doiprefix\url{10.1016/j.icarus.2017.05.026}
  (\bibinfo{year}{2018}).
\newblock \eprint{1706.02091}.

\bibitem{density}
\bibinfo{author}{{Krasinsky}, G.~A.}, \bibinfo{author}{{Pitjeva}, E.~V.},
  \bibinfo{author}{{Vasilyev}, M.~V.} \& \bibinfo{author}{{Yagudina}, E.~I.}
\newblock \bibinfo{journal}{\bibinfo{title}{{Hidden Mass in the Asteroid
  Belt}}}.
\newblock {\emph{\JournalTitle{Icarus}}} \textbf{\bibinfo{volume}{158}},
  \bibinfo{pages}{98--105}, \doiprefix\url{10.1006/icar.2002.6837}
  (\bibinfo{year}{2002}).

\bibitem{Ralph}
\bibinfo{author}{{Bacon}, R.~H.}
\newblock \bibinfo{journal}{\bibinfo{title}{{Logarithmic Spiral: An Ideal
  Trajectory for the Interplanetary Vehicle with Engines of Low Sustained
  Thrust}}}.
\newblock {\emph{\JournalTitle{American Journal of Physics}}}
  \textbf{\bibinfo{volume}{27}}, \bibinfo{pages}{164--165},
  \doiprefix\url{10.1119/1.1934788} (\bibinfo{year}{1959}).

\end{thebibliography}
\end{document}